\documentclass{aa}  
\usepackage{graphicx}
\usepackage{txfonts}
\usepackage{graphicx}
\usepackage{amsmath}
\usepackage[]{hyperref}
\usepackage{adjustbox}
\usepackage{enumitem}
\usepackage{natbib}
\usepackage{url}
\usepackage{times}
\def\spose#1{\hbox to 0pt{#1\hss}}
\newcommand\lsim{\mathrel{\spose{\lower 3pt\hbox{$\mathchar"218$}}
     \raise 2.0pt\hbox{$\mathchar"13C$}}}
\newcommand\gsim{\mathrel{\spose{\lower 3pt\hbox{$\mathchar"218$}}
     \raise 2.0pt\hbox{$\mathchar"13E$}}}

\makeatletter
\renewcommand*\aa@pageof{, page \thepage{} of \pageref*{LastPage}}
\makeatother

\begin{document} 

\title{Black hole mass and spin estimates of the most distant quasars}
	\titlerunning{Black hole mass and spin of the most distant quasars}
\author{Samuele Campitiello\inst{1}\thanks{\email{sam.campitiello@gmail.com}}
\and Annalisa Celotti\inst{1,2,3}
\and Gabriele Ghisellini\inst{2}
\and Tullia Sbarrato\inst{4}}

\institute{SISSA, Via Bonomea 265, I–34135, Trieste, Italy
   \and INAF - Osservatorio Astronomico di Brera, via E. Bianchi 46, I-23807, Merate, Italy
\and INFN – Sezione di Trieste, via Valerio 2, I-34127 Trieste, Italy \and
Dipartimento di Fisica "G. Occhialini", Università di Milano - Bicocca, Piazza della Scienza 3, I-20126 Milano, Italy
            }

\date{Received ; accepted }

\abstract{We investigate the properties of the most distant quasars ULASJ134208.10+092838.61 ($z = 7.54$), ULASJ112001.48+064124.3 ($z = 7.08$) and DELSJ003836.10-152723.6 ($z = 7.02$) studying their Optical-UV emission that shows clear evidence of the presence of an accretion disk. We model such emission applying the relativistic disk models KERRBB and SLIMBH for which we have derived some analytical approximations to describe the observed emission as a function of the black hole mass, accretion rate, spin and the viewing angle. We found that: 1] our black hole mass estimates are compatible with the ones found using the virial argument but with a smaller uncertainty; 2] assuming that the virial argument is a reliable method to have a black hole mass measurement (with no systematic uncertainties involved), we found an upper limit for the black hole spin of the three sources: very high spin values are ruled out; 3] our Eddington ratio estimates are smaller than those found in previous studies by a factor $\sim 2$: all sources are found to be sub-Eddington. Using our results, we explore the parameter space (efficiency, accretion rate) to describe the possible evolution of the black hole assuming a $\sim 10^{2-4} M_{\odot}$ seed: if the black hole in these sources formed at redshift $z = 10 - 20$, we found that the accretion has to proceed at the Eddington rate with a radiative efficiency $\eta \sim 0.1$ in order to reach the observed masses in less than $\sim 0.7$ Gyr. }

\maketitle


\section{Introduction} \label{sec:intro}

The existence of super-massive black holes (SMBHs) at redshifts $z > 6$ has been confirmed by a growing number of observations (\citealt{Fanetal}; \citealt{Baretal}; \citealt{WilMcJa}). Since the Universe was less than $\sim 1$ billion years old, their large masses ($M \sim 10^8 - 10^{10} M_{\odot}$) represent one of the most challenging aspects of such objects and the issue related to their rapid growth is still under debate. Several authors proposed different evolutionary scenarios (e.g. \citealt{HaiLoe}; \citealt{VolRees}; \citealt{LiYetal}; \citealt{Pelup}; \citealt{TanHai}; \citealt{Li}). The two possible frameworks adopted to understand the evolution of SMBHs are:
\begin{itemize}[label=$\bullet$]
	\item Merging between multiple black holes (BHs) (e.g. \citealt{Voletal}); this could have led to an accelerated growth of the black hole mass. The production of gravitational waves is important (e.g. Fig. 2 in \citealt{Seo13}) and the possible recoil effects could have slowed the BH growth down and prevented it growing to sufficiently large masses (\citealt{Haiman}; \citealt{Ole}; \citealt{Vol07}) \footnote{\citet{Yoo} showed that the gravitational wave recoil problem can be overcome within certain conditions, making a BH grow quickly without invoking super-Eddington accretion.}. A large number of black hole mergings is required to form a $\sim 10^9 M_{\odot}$ in a short amount of time and this could happen in a hierarchical process (\citealt{VolNat09}; \citealt{Seo13}).
	\item Accretion of matter onto the BH (e.g. \citealt{RusBeg}; \citealt{Koushetal}; \citealt{VolRees}; \citealt{Dottietal}); this process can happen in two different ways (or a combination of the two):
	\begin{itemize}
		\item Chaotic accretion of `blobs' of matter: this could have led to fast super-Eddington accretion with a rapid growth of the BH mass. 			
		\item Accretion through a disk-like structure: this scenario leads to the production of the observed thermal UV emission\footnote{The origin of the so-called “big blue bump” emission in Active Galactic Nuclei (AGN) is still under debate and, as discussed by, for example \citet{Kora}, there are some significant inconsistencies between the predictions of the standard disk models and observations: the broadband continuum slopes at optical/near-UV wavelengths (e.g., \citealt{Neugetal}; \citealt{Berketal}; \citealt{DavWooBla}); X-rays and the soft X-ray excess (e.g. \citealt{Pounds}; \citealt{NanPou}); micro-lensing observations of the accretion disk size (e.g. \citealt{RauchBland}).}. Several authors (e.g. \citealt{SS}, hereafter SS; \citealt{NovTho}; \citealt{LaoNet}; \citealt{CzZb}; \citealt{Hub2000}) described this emission whose features depend on different parameters like the accretion rate, the BH mass and its spin. 
	\end{itemize}
\end{itemize}

Both merging and accretion have an effect on the BH spin. In a BH-BH merging or a chaotic accretion scenario, objects falling onto a central BH from different directions affect its spin amplitude and orientation (e.g. \citealt{Dottietal}): if this happens randomly, the expected adimensional spin value is $a \sim 0$. If the accretion occurs through a disk, the spin orientation is nearly constant and the BH eventually spins up to its maximum value after roughly doubling its mass (\citealt{Bardeen}; \citealt{Thorn74}).\footnote{If the counteracting torque produced by the radiation emitted by the disk is taken into account, the 'canonical' equilibrium value for the adimensional black hole spin is $a = 0.9982$ (\citealt{Thorn74}).} This suggests that if a disk-like structure, producing the observed UV bump, is present around a BH with a coherent angular momentum for a sufficiently long time, the spin must be large. Other effects may spin the BH down, like the formation of relativistic jets through the Blandford–Znajek process (\citealt{BlandZna}): nevertheless some authors showed that even in the presence of such process, the BH spins rapidly with $a > 0.7$ (e.g. \citealt{Luetal}; \citealt{Wang}).

If accretion occurred mainly through a disk, a BH could have reached the maximum spin value in the early stage of its evolution, but this scenario alone could explain the presence of SMBHs at high redshifts only if the BH seed mass is $\gsim 10^7 M_{\odot}$ and accreting at the Eddington rate (\citealt{Li}). This is in contrast with what has been proposed by several authors (\citealt{VolRees}; \citealt{Vol10}; \citealt{AlHi}) who suggested a smaller seed mass of the order of $\sim 10^2 - 10^5 M_{\odot}$. In the latter case, super-Eddington accretion represents a solution for the rapid growth problem. In this context, \citet{Lapietal14} showed that a BH can grow by accretion in a self-regulated regime with radiative power that can slightly exceed the Eddington limit at high redshifts: in this scenario, the radiative efficiency of the disk is $\eta = 0.15$, large enough to produce a significant luminosity. 

In view of this discussion, the aim of our work is to estimate disk luminosity and BH mass, and to constrain the spin of the highest redshift quasars. Specifically, we focused on the three highest redshift objects known up to the end of 2018: ULASJ134208.10+092838.61 (J1342) at $z=7.54$ (\citealt{Banados17}, hereafter BN18), the second most distant quasar ULASJ112001.48+064124.3 (J1120) at $z = 7.08$, studied by \citet{Mor11} (hereafter MR11), and the recent discovery DELSJ003836.10-152723.6 (J0038) at redshift $z=7.02$ (\citealt{Wangetal18}, hereafter W18). The first quasar (QSO) is a source with a radio-loudness $\mathcal{R} = 12.4$ (\citealt{Venemans})\footnote{The radio-loudness is defined as $\mathcal{R} = S_{\rm 5\ GHz,\ rest} / S_{\rm 4400\ \text{\AA},\ rest}$, where $S$ is the flux density at a specific rest-frame frequency.}, at the border of the radio-loud/quiet divide ($\mathcal{R} > 10$); for the second one $\mathcal{R} < 0.5 - 4.3$ has been estimated, depending on the assumed radio spectral index (\citealt{Momj})\footnote{They defined the radio-loudness as $\mathcal{R} = L_{\nu,\ \rm 1.4\ GHz,\ rest} / L_{\nu,\ \rm 4400\ \text{\AA},\ rest}$, where $L_{\nu}$ is the luminosity density at a specific rest-frame frequency.}; no radio detections are available for the third source. These sources can be classified as Type 1 QSOs, and as such we do not expect them to be strongly absorbed (by a dusty torus).

In order to estimate luminosity, mass and spin, we 'fit' their spectral energy distribution (SED) with an accretion disk model. We call this procedure SED-fitting method. This method is widely used (e.g. \citealt{Francis}; \citealt{Molendi}; \citealt{LaoDav_}; \citealt{Caldero}; \citealt{Sbaretal13}; \citealt{CastiDe}; \citealt{Capel1,Capel3}; \citealt{GhiTav15}; \citealt{Cast}; \citealt{Sbaretal}; \citealt{Maj}); some of these works show a good agreement between the SED-fitting and the single epoch virial method for deriving the BH mass (e.g. \citealt{Castietal}; \citealt{Cast}; \citealt{Maj}) and that motivated us to study the three sources using (relativistic) accretion disk models\footnote{The accretion disk model reliability to infer the BH mass can be verified by using BH mass estimates from the reverberation mapping (RM) technique. However, in this case a large sample of sources with good optical-UV data is needed and the uncertainties related to the scale factor $f$, not related to the quality of data (\citealt{BenKa}), must be taken into account.}.

First we use the relativistic model KERRBB (\citealt{Lietal}) developed to describe the radiation emitted by a thin disk around a stellar Kerr BH (\citealt{Campiti} - hereafter C18 - found analytic expressions practical also for SMBHs): one of the main assumptions of this model is that all the heat generated in the disk is immediately radiated away and this assumption is physically consistent only for low accretion rates (i.e. the Eddington ratio must be $\lambda \lsim 0.3$; see \citealt{LaoNet}; \citealt{McClint}, but larger than $\lambda \sim 0.01$, since for smaller accretion rates the advection-dominated accretion flow regime sets in, changing the main properties of the accretion disk; \citealt{Ichi}; \citealt{Reesetal82}; \citealt{NarMC}). For higher accretion rate values, advection dominates over radiative energy transport in the inner parts of the disk and this latter is thought to inflate in the so-called ``slim'' regime (e.g. \citealt{Abretal}): for this reason, supposing that high-redshift quasars have gone through a phase with a large accretion rate to attain a fast growth, we compared the KERRBB results with the prediction of the slim disk model SLIMBH (i.e. \citealt{Sad09}; \citealt{SadwAbra09}; \citealt{SadwAbra}; \citealt{StraubDo}) \footnote{Both these models are implemented in the interactive X-ray spectral fitting program XSPEC (see \citealt{ArnaudXPS} and references therein).} whose application could be more appropriate for near and slightly super-Eddington AGNs (see \citealt{Kora}). Also this model is designed for stellar mass BHs: we extended its usage for SMBHs (following the same analysis done by C18 for KERRBB) deriving some analytical approximation to describe the observed disk emission as a function of the BH mass, spin, Eddington ratio and viewing angle (see Appendix \ref{APP.B}). 

\begin{figure}
\centering
\hskip -0.2 cm
\includegraphics[width=0.49\textwidth]{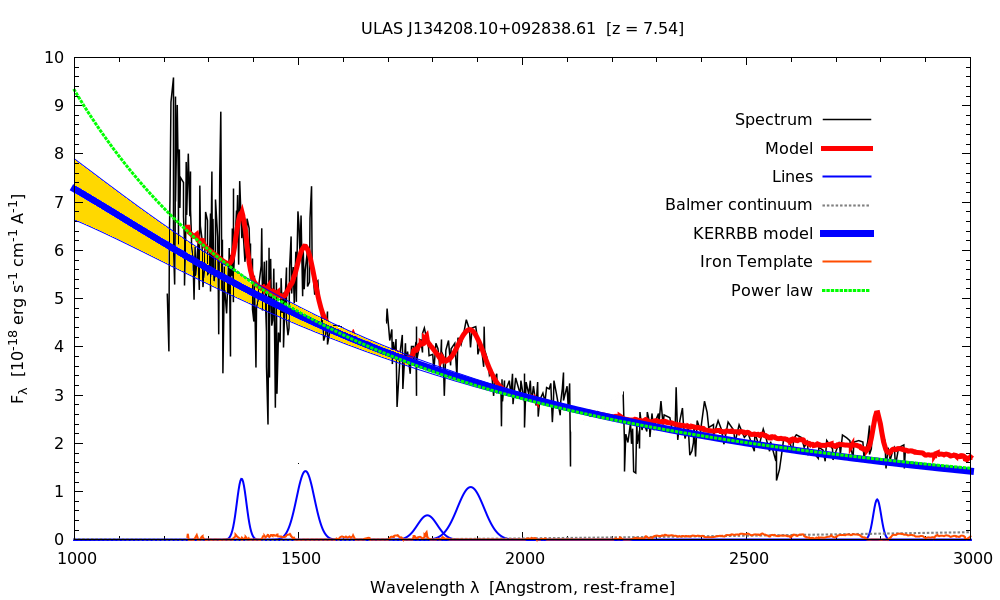} \includegraphics[width=0.49\textwidth]{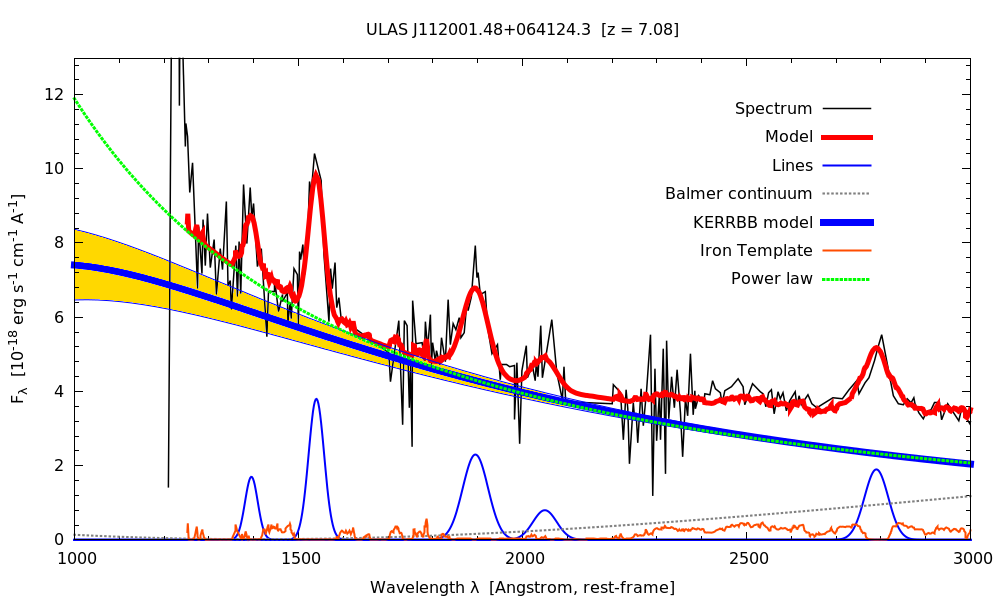} \includegraphics[width=0.49\textwidth]{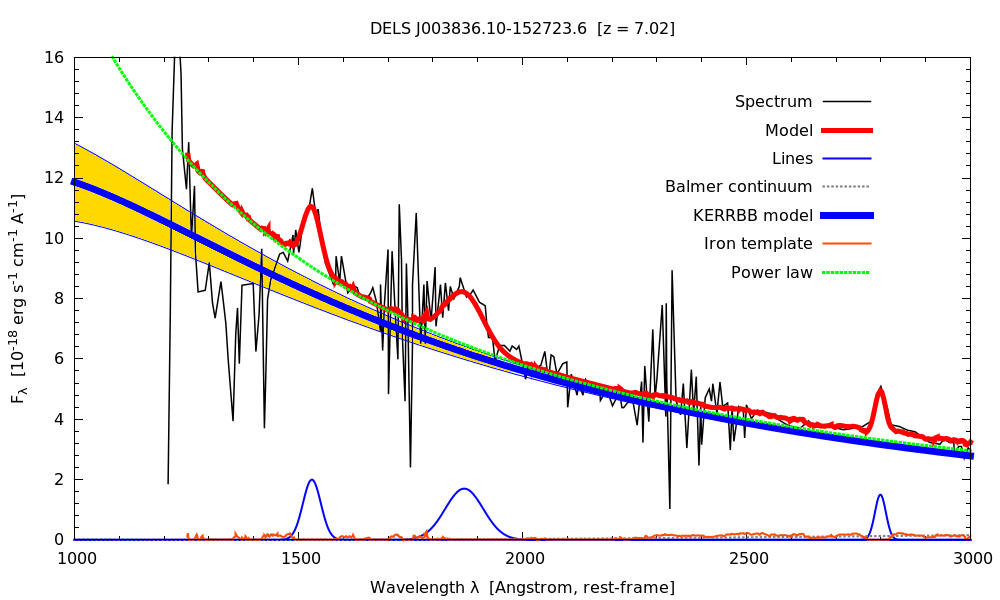}
\caption{Spectrum fit of the three sources as a power-law continuum (dotted green line) plus the iron complex (orange line), the Balmer continuum (dotted gray line) and some prominent emission lines (thin blue line) like MgII, CIII, CIV and SiIV. The red line is the model of the spectrum given by the sum of all these components. The thick blue line is the KERRBB model of the continuum for which we defined a confidence interval (yellow area): the model overlaps well the power law for $\lambda > 1700\ \text{\AA}$ (the difference at shorter wavelengths is due to the fact that the accretion disk model is not a power law around the spectral peak). Regions between the J-H and H-K$_{\rm s}$ bands are affected by low sky transparencies (these regions are not shown in the J1342 spectrum -- BN18; for the other two sources they are visible as very noisy features): for this reason we did not include them in the fitting procedure.} 
\label{SED1}
\end{figure}

In Sect. \ref{sec:2} we describe the procedure adopted for the 'fitting', discussing the possible uncertainties in the observed spectra due to various sources of absorption and emission, and estimating the consequent uncertainties on the resulting physical parameters of the fitted models. We then present the results of the fits in terms of luminosities (Sect. \ref{2.1}), BH masses (Sect. \ref{2.2}), spins (Sect. \ref{2.3}) and Eddington ratios (Sect. \ref{2.4}) and compare the findings of the Kerr versus slim disk modelling (Sect. \ref{2.5}).
In Sect. \ref{sec:3}, we discuss the possible BH evolution with different assumptions on the radiative efficiency and accretion rate. A discussion and the final conclusions are the content of Sect. \ref{sec-concl}.

In this work, we adopt a flat cosmology with $H_0=68$ km s$^{-1}$ Mpc$^{-1}$ and $\Omega_{\rm M}=0.3$ (Planck Collaboration XIII, 2015). 


\section{Spectral energy distribution fitting: Results} \label{sec:2}

\begin{figure}
\begin{center}
\hskip -0.2 cm
\includegraphics[width=0.42\textwidth]{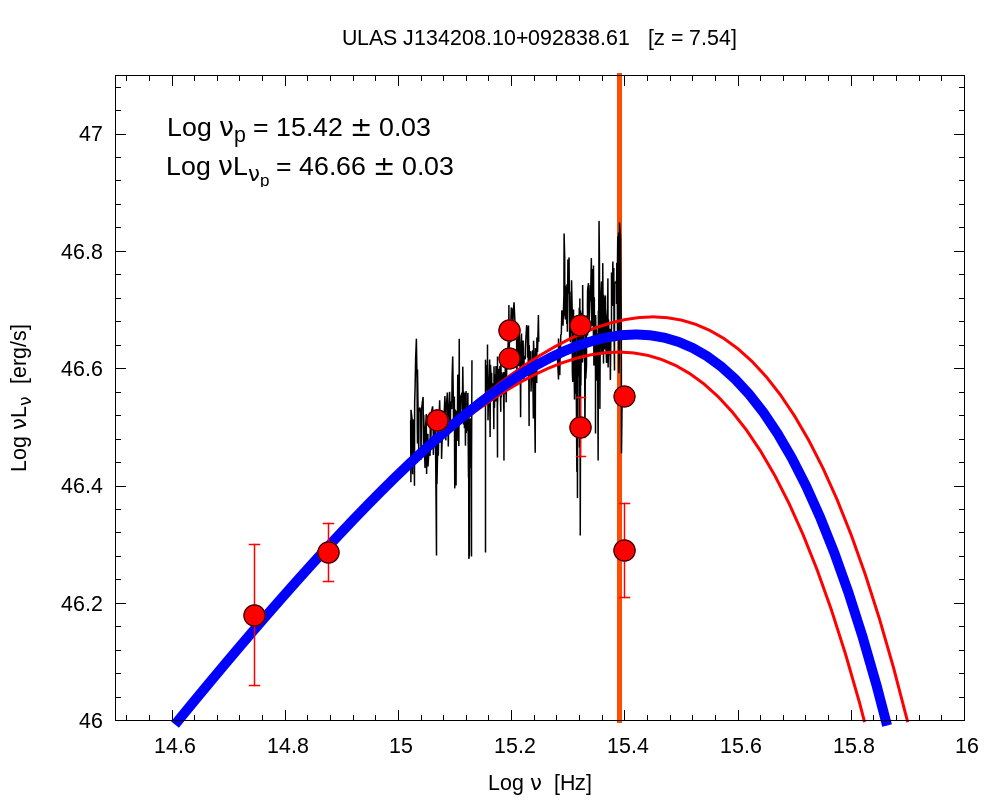}  \includegraphics[width=0.42\textwidth]{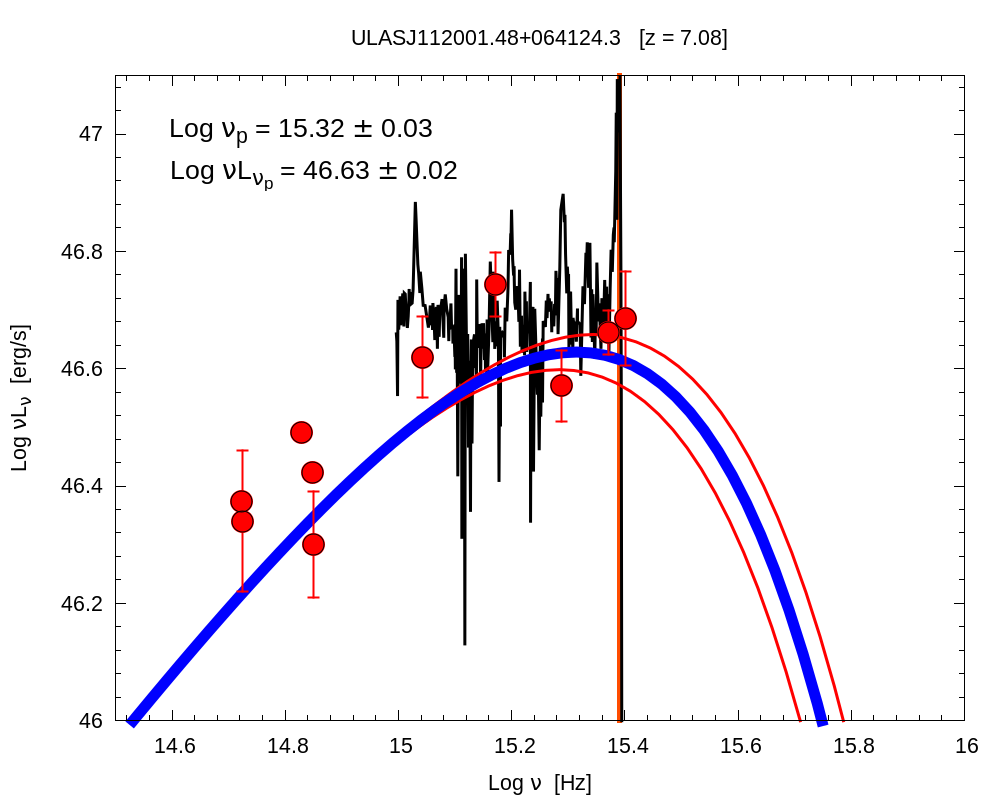}
\includegraphics[width=0.42\textwidth]{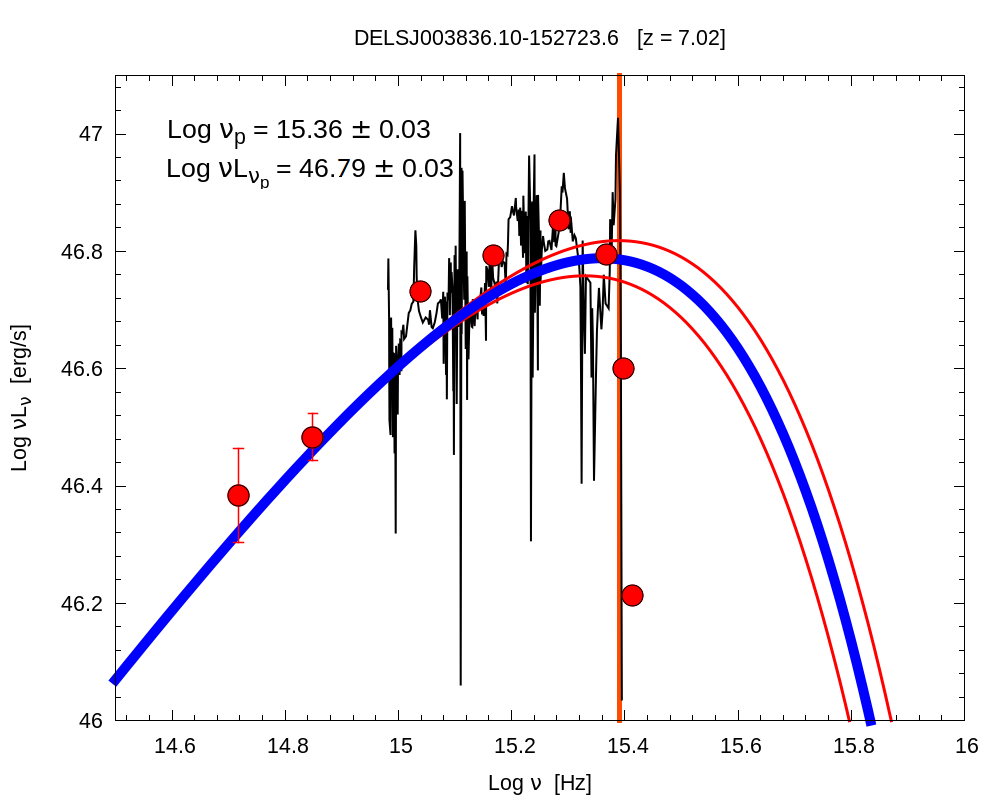}
\caption{Top panel: SED of the quasar J1342 in the rest frame whose observed disk luminosity is $L^{\rm obs}_{\rm d} = (9.8 \pm 0.7) \cdot 10^{46}$ erg s$^{-1}$. Photometric data (red points) and spectrum (black line) are from \citet{Banados17} (see this work for details regarding these data). Central panel: SED of the quasar J1120 in the rest frame with an observed disk luminosity $L^{\rm obs}_{\rm d} = (9.1 \pm 0.5) \cdot 10^{46}$ erg s$^{-1}$. Photometric data (red points) and spectrum (black line) are from the work by \citet{Mor11}. Bottom panel: SED of the quasar J0038 in the rest frame with an observed disk luminosity $L^{\rm obs}_{\rm d} = (1.3 \pm 0.7) \cdot 10^{47}$ erg/s. Photometric data (red points) and spectrum (black line) are from the work by \citet{Wangetal18}. In these panels the blue curve is the 'best fit' obtained using the relativistic model KERRBB. The red curves describe the confidence interval for the spectrum peak frequency and luminosity from the model. The vertical orange line indicates the Ly$\alpha$ line frequency: at frequencies larger than this latter, data are strongly affected by intervening Ly$\alpha$ clouds along our line of sight and therefore we did not include these data in our fit.} 
\label{SED}
\end{center}
\end{figure}

In this section, we fit the optical-UV SED of the sources J1342, J1120 and J0038 ($\lambda - F_{\lambda}$ plot in Fig. \ref{SED1}; $\nu - \nu L_{\nu}$ plot in Fig. \ref{SED}), in order to extrapolate information about the observed disk luminosity, the BH mass, spin and Eddington ratio; then we compare the results with previous estimates. In order to infer these quantities, we interpreted the optical-UV "bump" as the emission produced by an accretion disk around a SMBH: by adapting two relativistic models (first KERRBB and then SLIMBH) to the spectrum (black line in Figs. \ref{SED1}-\ref{SED}) we estimated the peak frequency $\nu_{\rm p}$ and luminosity $\nu_{\rm p} L_{\nu_{\rm p}}$ needed to infer all the information (see Appendix). We have associated to these last quantities an uncertainty ($\pm 0.03$ dex) which defines a confidence interval (yellow area in Fig. \ref{SED1} and red curves in Fig. \ref{SED}). In the fit procedure we did not include photometric data (red points) and the associated uncertainties because they may be contaminated by lines emission (e.g. Ly$\alpha$ and MgII, for J1120) and/or affected by absorption (e.g. points at Log $\nu \gsim 15.4$ for J1342 and J0038). 

We limited our fitting to the spectral range $\lambda \sim 1800-3000\ \text{\AA}$ where the uncertainties on the underlying continuum are minimized. The confidence interval of the modelling just mentioned is thus determined by the range of predicted spectra which does not alter the spectrum in such a spectral range. It is relevant to stress that the location of the spectrum peak can be estimated even if it is slightly outside the frequency range covered by data (i.e. at larger frequencies). This is because we can fit the changing curvature of the spectrum when it is approaching its peak.  

In the following we discuss in some detail the effects that different phenomena (which could modify the UV spectrum) could have on the goodness of the modeling.

\begin{itemize}
	\item Intrinsic Dust Extinction: Figures \ref{SED1} and \ref{SED} show the SED of the three sources without any correction from possible dust absorption: since these sources are Type 1 QSOs, we do not expect any (strong) dust absorption of the accretion disk emission from the surrounding molecular torus; the latter is thought to have an average opening angle of $\sim 45^{\circ}$ from the normal to the disk (see e.g. \citealt{Caldero12}). For this reason we assumed an observation of these sources with a viewing angle $\theta_{\rm v} < 45^{\circ}$ from the normal to its accretion disk. However dust extinction could be present at a relatively low level. If corrected for dust, the spectrum becomes harder with a peak shifted at larger frequencies, increasing the disk luminosity, lowering the BH mass and increasing the Eddington ratio estimates. We checked the effect on our results by considering the extinction curves derived by \citet{Czetal}. A self-consistency requirement for modeling the spectrum as disk emission is that the slope of the corrected, de-reddened spectrum has to be softer than the theoretical value $F_{\nu} \propto \nu^{1/3}$. This translates into an upper limit for the extinction $E[B-V] \lsim 0.15$. Adopting this "extreme" correction, the resulting disk luminosity, BH mass and Eddington ratio would change by less than a factor $\lsim 2$.
	\item Dust Emission: This emission is mainly produced in the IR band, characterized by two significant bumps (silicate dust emission at around $\nu \sim 3 \cdot 10^{13}$ Hz and a hot component at around $\nu \sim 10^{14}$ Hz; \citealt{Barva}; \citealt{PierKro}; \citealt{Mor}; \citealt{HonKish}). Therefore, the contribution by such components is not expected to affect the spectral region of the disk emission fit.
	\item Contribution from lines: We fitted the spectrum of the three sources (Fig. \ref{SED1}) as a power-law continuum plus the iron complex (\citealt{VesterWi}; \citealt{Tsuzu}), the Balmer continuum (e.g., \citealt{Derosa}) and prominent emission lines (modeling them with a Gaussian, e.g. MgII, CIII, CIV, SiIV), using the fitting routine implemented in {\rm GNUPLOT} (non-linear least-squares Marquardt-Levenberg algorithm). Then we performed the same procedure using the KERRBB model to describe the AGN continuum instead of the power law: as shown in Fig. \ref{SED1}, the KERRBB model overlaps well the power law for $\lambda > 1700\ \text{\AA}$ (the difference at shorter wavelengths is due to the fact that the accretion disk model is not a power law around the spectral peak). Therefore, when the contributions from lines are subtracted from the total spectrum (and even if uncertainties on the spectrum itself are taken into account)\footnote{The uncertainties are shown in Figure 1 of BN18 (gray line), MR11 (black line) and W18 (gray line): they are rather small at 1$\sigma$ level (the percentage uncertainty is less than $\sim 20 \%$).}, the continuum emission is well reproduced by our KERRBB model inside our confidence interval (yellow area in Fig. \ref{SED1}). 
	\item Poor sky transparencies: Regions between the J-H and H-K$_{\rm s}$ bands are affected by low sky transparencies (Figs. \ref{SED1}-\ref{SED}). For J1342 these regions are not shown in the original spectrum (\citealt{Banados17}); instead, for the other two sources, these regions are visible but they are very noisy. Therefore we did not include them in the fitting procedure. Even if they were included, the fit would not change significantly because the best model is well determined by the rest of the spectrum.
	\item IGM absorption: We considered the correction of the spectrum from the intergalactic medium (IGM) absorption (at least for the source J1342), following the reconstruction of the AGN intrinsic emission done by BN18: such correction affects only the Ly$\alpha$ line emission band without modifying the spectrum at lower frequencies, and so without affecting our best fit.
	\item Broad Absorption Lines: J0038 shows some absorption troughs that may have caused the high frequency part of the spectrum to be dimmer. The spectral analysis showed that the fit is not influenced by the absorption features of the spectrum; moreover, our confidence interval is in good agreement with the quasar composite spectrum constructed with $\sim 200$ SDSS quasars, compared with that of J0038 and shown in Fig. 1 of W18.
\end{itemize}

We conclude that, although the above effects can alter the spectrum, the uncertainties are mostly consistent with the estimated confidence interval of the model (corresponding to the yellow area in Fig. \ref{SED1} and the two red curves reported in Fig. \ref{SED}). The major uncertainty on the spectrum peak position estimates could come from dust absorption, affecting the derived disk luminosity, BH mass and Eddington ratio at most by a factor $\lsim 2$.

\subsection{Observed disk luminosity}\label{2.1}

The observed disk luminosity can be estimated from the spectrum peak luminosity, $L^{\rm obs}_{\rm d} = \int L_{\nu} d \nu \sim 2\ \nu_{\rm p} L_{\nu_{\rm p}}$.\footnote{A similar relation has been found by \citet{Caldero} using the classical SS model. See also C18.} We found $L^{\rm obs}_{\rm d} = (9.8 \pm 0.7) \cdot 10^{46}$ erg s$^{-1}$ for J1342; $L^{\rm obs}_{\rm d} = (9.1 \pm 0.5) \cdot 10^{46}$ erg s$^{-1}$ for J1120; and $L^{\rm obs}_{\rm d} = (1.3 \pm 0.7) \cdot 10^{47}$ erg s$^{-1}$ for J0038. These values are smaller than the bolometric luminosities estimated by BN18, MR11 and W18 because their results are based on the bolometric correction $L_{\rm bol} = C \times L_{\rm 3000}$ (where $L_{\rm 3000}$ is the luminosity at 3000 $\text{\AA}$) \footnote{These bolometric estimates also take into account the luminosity contribution in the IR and X-ray bands and the bolometric correction factor $C$ has an uncertainty due to the scatter in the SEDs for individual quasars (e.g., \citealt{Rich06}; \citealt{Vasu}).} that overestimates the disk luminosity by a factor $\sim 2$ (\citealt{Caldero}). It is important to point out that the observed disk luminosity is different with respect to the total disk luminosity $L_{\rm d} = \eta \dot{M} c^2$ (spin dependent) by a factor $2 \cos \theta_{\rm v}$ if one considers a SS model (\citealt{Caldero}), or by a factor depending on the viewing angle and the BH spin if one considers the KERRBB/SLIMBH model (see C18). 

\begin{figure}
\begin{center}
\hskip -0.2 cm
\includegraphics[width=0.405\textwidth]{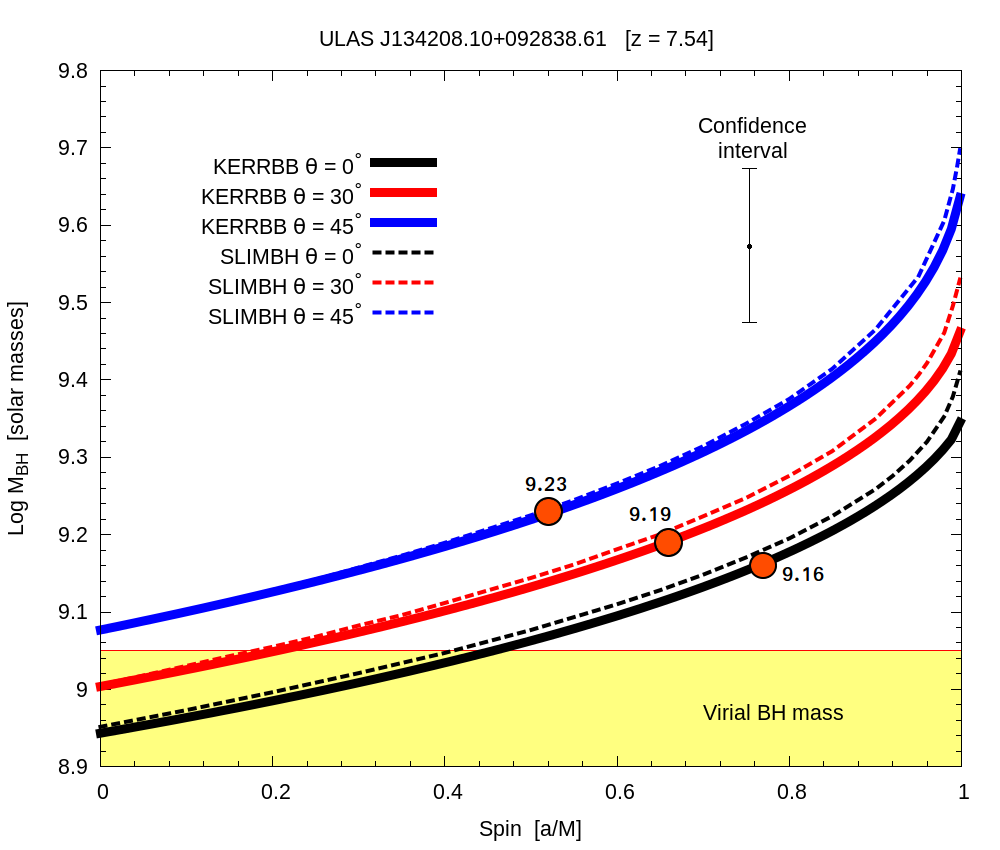} 
\includegraphics[width=0.41\textwidth]{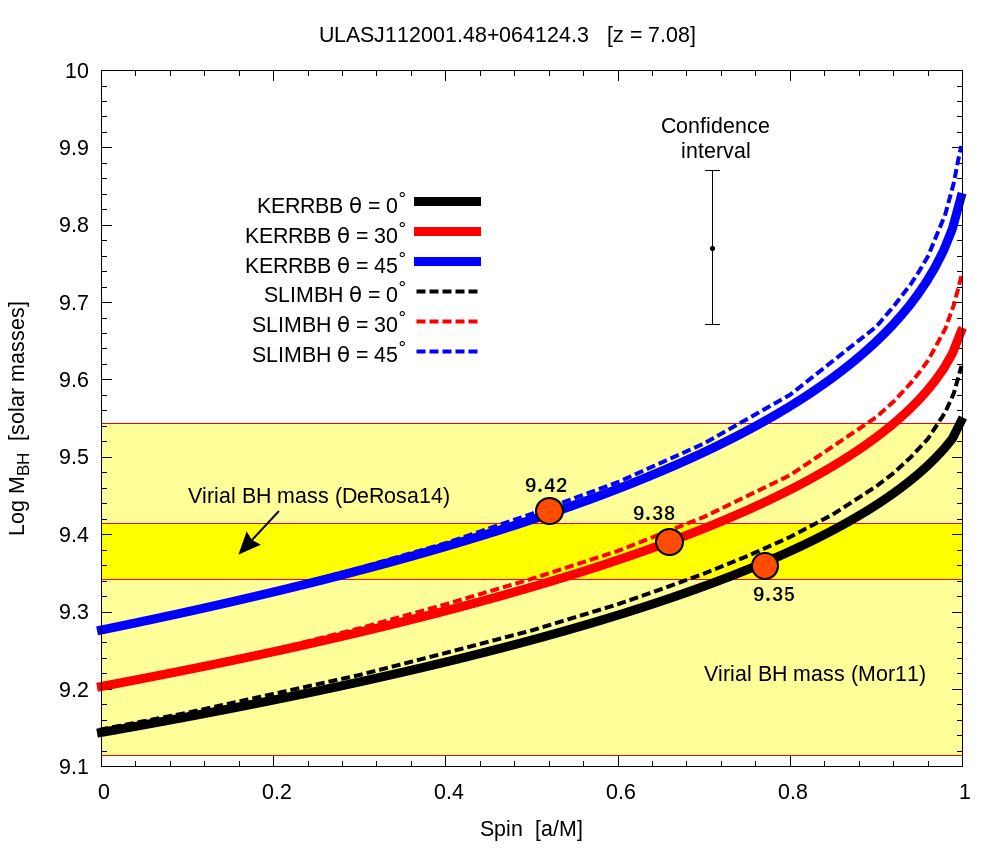}
\includegraphics[width=0.41\textwidth]{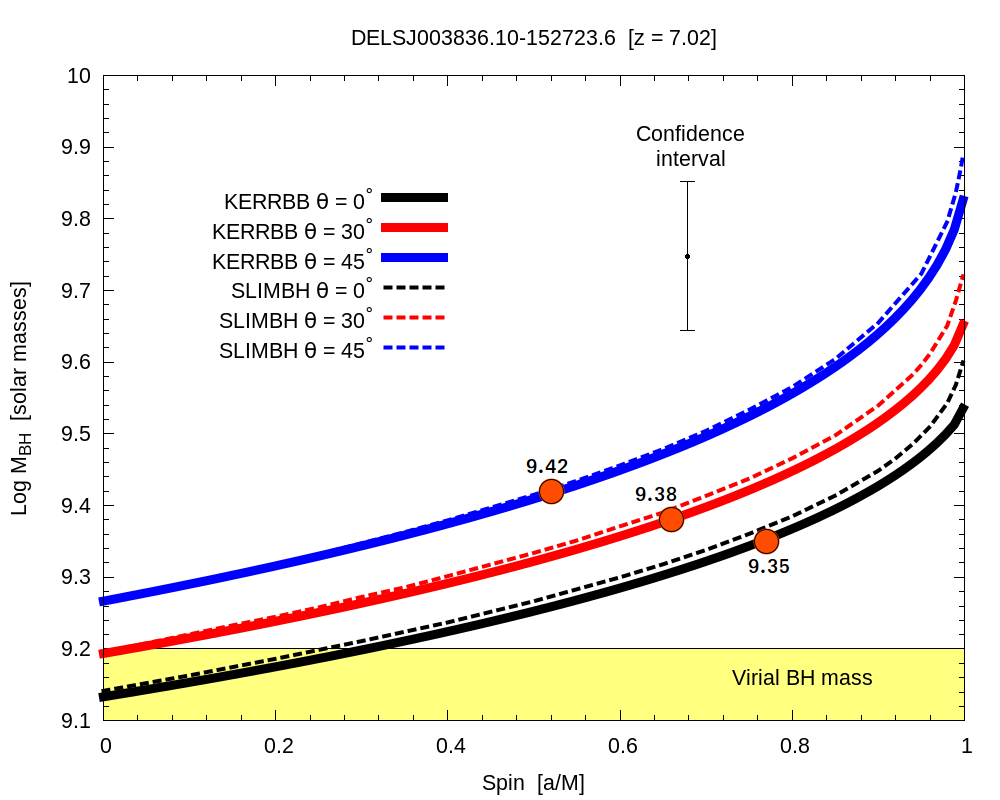} 
\caption{Top panel: BH mass as a function of the BH spin computed using the KERRBB and SLIMBH models (solid and dashed lines, respectively) for a fixed spectrum peak position and different viewing angles ($0^{\circ}$, $30^{\circ}$, $45^{\circ}$) for the source J1342. The yellow area represents the virial mass range (Log $M_{\rm vir}/M_{\odot} = 8.89^{+0.16}_{-0.12}$) estimated by BN18 using the MgII line. Central panel: same comparison for the source J1120. The light yellow shaded area represents the virial mass range (Log $M_{\rm vir}/M_{\odot} = 9.30^{+0.24}_{-0.19}$) estimated by MR11 using the MgII line; the dark yellow area is the virial mass range (Log $M_{\rm vir}/M_{\odot} = 9.38^{+0.03}_{-0.04}$) estimated by \citet{Derosa} using the same procedure. Bottom panel: same comparison for J0038. The yellow shaded area is the virial mass (Log $M_{\rm vir}/M_{\odot} = 9.12^{+0.08}_{-0.09}$) estimated by W18 using the MgII line. Orange dots represent the BH mass solutions coming from the classical SS model: these correspond to particular KERRBB/SLIMBH solutions with a precise spin value. The confidence interval ($\pm 0.1$ dex) is derived assuming a small uncertainty on the spectrum peak position on which the BH mass estimates are based.} 
\label{KERRBBmass}
\end{center}
\end{figure}

\subsection{Black hole mass estimates}\label{2.2}

Figure \ref{KERRBBmass} shows the KERRBB BH mass (solid lines) for the three sources as a function of the spin, for $\theta_{\rm v} = 0^{\circ} - 30^{\circ} - 45^{\circ}$ (following C18, we used the same procedure to find analytic expressions for different viewing angles; see Appendix \ref{APP.A}). These solutions describe the same spectrum with the same peak position: as shown by C18, if the spectrum peak is fixed, the KERRBB model is degenerate in mass $M$, accretion rate $\dot{M}$, and spin $a$.

In fact, assume that a spectrum can be described for some values of $M$, $\dot{M}$ and $a$: if the spin value is increased, the inner radius of the disk moves to closer orbits. As a consequence, the radiative efficiency $\eta$ increases, making the disk luminosity increase as well. Consequently the spectrum shifts to higher luminosities and frequencies. However by increasing $M$ (i.e. moving the spectral peak to lower $\nu$) and decreasing $\dot{M}$ (i.e. shifting the spectrum to lower $\nu L_{\nu}$) conveniently, it is possible to reproduce the same original spectrum (see Fig. 4 in C18). 

For illustration, in Fig. \ref{KERRBBmass} we show the values of the BH mass corresponding to the fitting with a classical SS model (orange dots), that is, with no spin no relativistic corrections: this model can mimic a KERRBB model with the same $M$, $\dot{M}$ and $\theta_{\rm v}$, and a specific spin $a$ (see C18 for more details). Our estimates of the BH mass of the three sources for $a=0 - 0.9982$ are summarized in Table \ref{ap:alltabresres}. In the same figure, the single epoch virial mass ranges (from the Mg II line) as estimated by BN18, MR11, \citet{Derosa} and W18 are shown as yellow areas. Our results are compatible with the virial mass estimates but with typically smaller uncertainties compared to the systematic $\sim 0.5$ dex in the local scaling relations for virial estimates (\citealt{VesterOs}).

\subsection{Black hole spin estimates}\label{2.3}

Assuming that the virial estimates are reliable measurements of BH masses (with no systematic uncertainties involved), the overlapping between the yellow area (indicating the virial BH mass range) and the KERRBB solutions in Fig. \ref{KERRBBmass} can in principle be used to find some constraints on the BH spin: taking the reported quantities for the source J1342, the BH spin would have to be $a < 0.5$. For J1120, using the virial estimates of MR11, an upper limit for the BH spin would be present only for $\theta_{\rm v} > 0^{\circ}$ ($a<0.9$ for $\theta_{\rm v} = 30^{\circ}$, $a<0.75$ for $\theta_{\rm v} = 45^{\circ}$); using the virial mass of DR14, the spin would be constrained in the range $0.25 < a < 0.85$. For J0038, the BH spin has to be $a<0.3$. These ranges would rule out the maximum spin solution, but clearly no systematic uncertainties on the virial BH mass estimates (major uncertainty) and on the spectrum peak position have been taken into account: if these are considered, no constrains on the spin can be derived.

\begin{figure}
\centering
\hskip -0.2 cm
\includegraphics[width=0.425\textwidth]{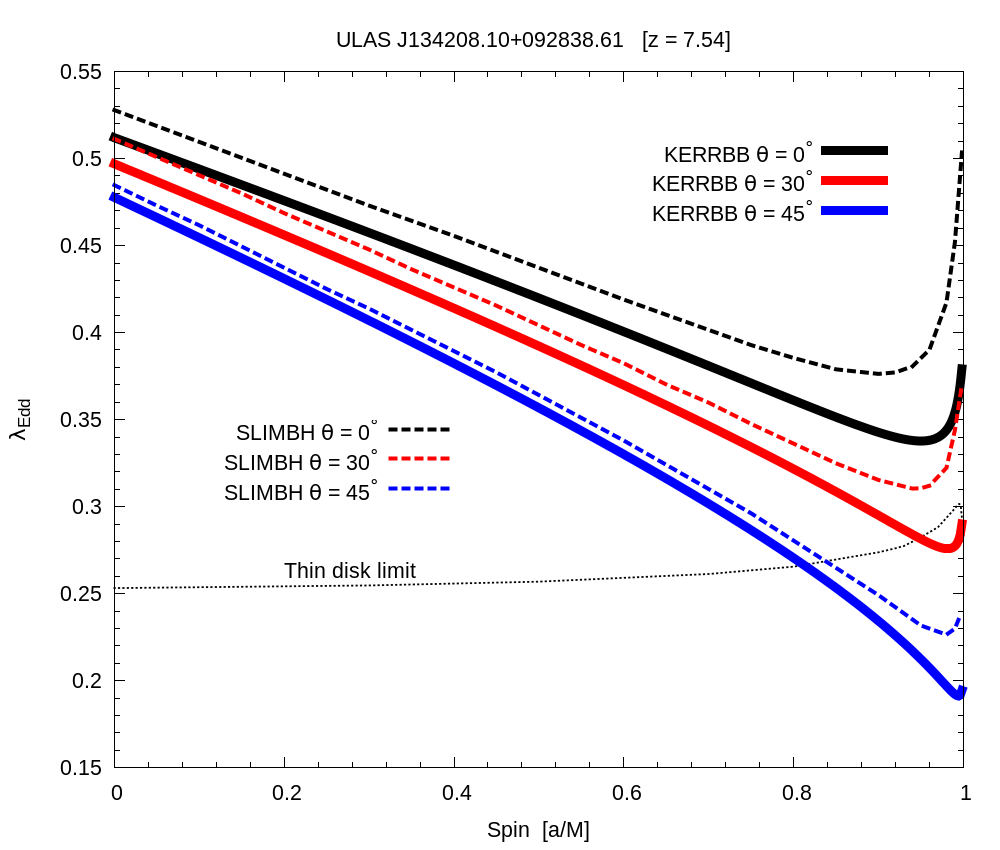} \includegraphics[width=0.425\textwidth]{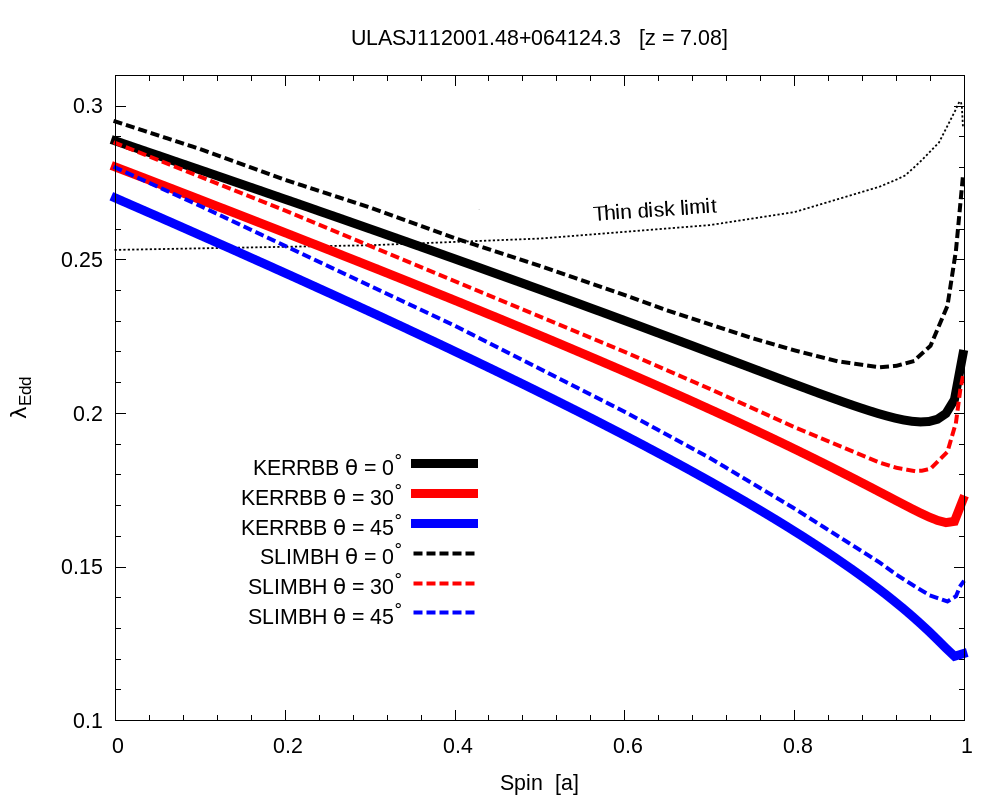} 
\includegraphics[width=0.425\textwidth]{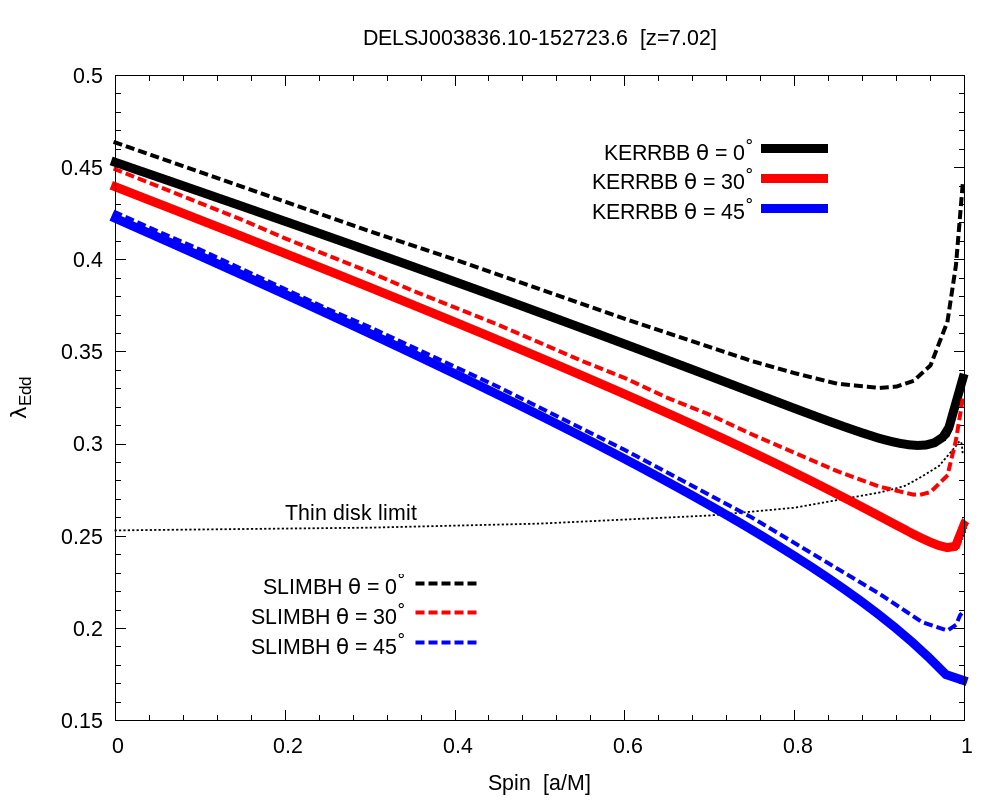}
\caption{Top panel: Eddington ratio $\lambda = L_{\rm d} / L_{\rm Edd}$ as a function of the BH spin $a$ for different viewing angles ($\theta_{\rm v} = 0^{\circ}$ black, $30^{\circ}$ red, $45^{\circ}$ blue) computed with the KERRBB (solid lines) and the SLIMBH (dashed lines) models for the source J1342. Central and bottom panels: same plot for the sources J1120 and J0038. The black dotted line is the thin disk limit for $\lambda$ following \citet{LaoNet}: the KERRBB solutions above this limit are not consistent with the thin disk approximation and another model (e.g. slim disk) must be used.} 
\label{KERRBBmass4}
\end{figure}

\begin{table*} 
\centering
\footnotesize
\begin{tabular}{lllllllllll}
\hline\\
\vspace{1mm}
Model & Source & $\theta_{\rm v}$ & $M_{\rm 0}$ & $\lambda_{\rm 0}$ & $M_{\rm 1}$ & $\lambda_{\rm 1}$ \\ 
\hline   
\hline   \\
KERRBB & J1342 & $0^{\circ}$ & $8.94 \pm 0.10 \hspace{2mm}$ (9.16) & $0.51 \pm 0.10 \hspace{2mm}$ (0.26) & $9.34 \pm 0.10$ & $0.38 \pm 0.10$ \\
 &  & $30^{\circ}$ & $9.00 \pm 0.10 \hspace{2mm}$ (9.19) & $0.49 \pm 0.10 \hspace{2mm}$ (0.28) & $9.46 \pm 0.10$ & $0.29 \pm 0.10$ \\
\vspace{1mm}
 &  & $45^{\circ}$ & $9.07 \pm 0.10 \hspace{2mm}$ (9.23) & $0.48 \pm 0.10 \hspace{2mm}$ (0.31) & $9.64 \pm 0.10$ & $0.19 \pm 0.10$ \\ 
 & J1120 & $0^{\circ}$ & $9.13 \pm 0.10 \hspace{2mm}$ (9.35) & $0.29 \pm 0.10 \hspace{2mm}$ (0.16) & $9.53 \pm 0.10$ & $0.22 \pm 0.10$ & \\ 
 &  & $30^{\circ}$ & $9.19 \pm 0.10 \hspace{2mm}$ (9.38) & $0.28 \pm 0.10 \hspace{2mm}$ (0.17) & $9.65 \pm 0.10$ & $0.17 \pm 0.10$ & \\
 &  & $45^{\circ}$ & $9.26 \pm 0.10 \hspace{2mm}$ (9.42) & $0.27 \pm 0.10 \hspace{2mm}$ (0.19) & $9.82 \pm 0.10$ & $0.13 \pm 0.10 \vspace{2mm}$ \\
 & J0038 & $0^{\circ}$ & $9.13 \pm 0.10 \hspace{2mm}$ (9.35) & $0.45 \pm 0.10 \hspace{2mm}$ (0.23) & $9.53 \pm 0.10$ & $0.34 \pm 0.10$ & \\ 
 &  & $30^{\circ}$ & $9.19 \pm 0.10 \hspace{2mm}$ (9.38) & $0.44 \pm 0.10 \hspace{2mm}$ (0.25) & $9.65 \pm 0.10$ & $0.26 \pm 0.10$ & \\
 &  & $45^{\circ}$ & $9.26 \pm 0.10 \hspace{2mm}$ (9.42) & $0.42 \pm 0.10 \hspace{2mm}$ (0.25) & $9.84 \pm 0.10$ & $0.17 \pm 0.10 \vspace{2mm}$ \\
\hline\\
SLIMBH & J1342 & $0^{\circ}$ & $8.95 \pm 0.10$ & $0.53 \pm 0.10$ & $9.41 \pm 0.10$ & $0.50 \pm 0.10$ \\
 &  & $30^{\circ}$ & $9.01 \pm 0.10$ & $0.51 \pm 0.10$ & $9.53 \pm 0.10$ & $0.37 \pm 0.10$ \\
\vspace{3mm}
 &  & $45^{\circ}$ & $9.08 \pm 0.10$ & $0.48 \pm 0.10$ & $9.70 \pm 0.10$ & $0.23 \pm 0.10$ \\ 
 & J1120 & $0^{\circ}$ & $9.14 \pm 0.10$ & $0.29 \pm 0.10$ & $9.60 \pm 0.10$ & $0.28 \pm 0.10$ & \\
 &  & $30^{\circ}$ & $9.19 \pm 0.10$ & $0.29 \pm 0.10$ & $9.72 \pm 0.10$ & $0.21 \pm 0.10$ & \\
 &  & $45^{\circ}$ & $9.26 \pm 0.10$ & $0.28 \pm 0.10$ & $9.89 \pm 0.10$ & $0.15 \pm 0.10 \vspace{2mm}$ \\
 & J0038 & $0^{\circ}$ & $9.14 \pm 0.10 \hspace{2mm}$ & $0.47 \pm 0.10 \hspace{2mm}$ & $9.60 \pm 0.10$ & $0.44 \pm 0.10$ & \\ 
 &  & $30^{\circ}$ & $9.20 \pm 0.10 \hspace{2mm}$ & $0.45 \pm 0.10 \hspace{2mm}$ & $9.72 \pm 0.10$ & $0.33 \pm 0.10$ & \\
 &  & $45^{\circ}$ & $9.27 \pm 0.10 \hspace{2mm}$ & $0.42 \pm 0.10 \hspace{2mm}$ & $9.89 \pm 0.10$ & $0.21 \pm 0.10 \vspace{2mm}$ \\
\hline
\hline   \\
\end{tabular}
\caption{Our estimates for the BH mass $M$ and Eddington ratio $\lambda$ of J1342, J1120 and J0038 considering spin $a=0$ ($M_{\rm 0}$, $\lambda_{\rm 0}$) and $a=0.9982$ ($M_{\rm 1}$, $\lambda_{\rm 1}$), and different viewing angles ($\theta_{\rm v} = 0^{\circ} - 30^{\circ} - 45^{\circ}$). The $\pm 0.10$ dex error comes from the small uncertainty on the spectrum peak position on which these estimates are based. For $a=0$, the two models give almost the same results (i.e. disk thickness and relativistic effects are negligible); for $a=0.9982$, the two models give different results by a factor of $\sim 1.2 - 1.3$. For comparison, we show in brackets also the SS results. \label{ap:alltabresres}}
\end{table*}

\subsection{Eddington ratios}\label{2.4}

Figure \ref{KERRBBmass4} shows the Eddington ratios of the sources computed for the different KERRBB solutions (solid lines) as a function of the spin. The Eddington ratio is defined as $\lambda = L_{\rm d} / L_{\rm Edd}$ (where $L_{\rm Edd} = \mathcal{K} (M/M_{\odot})$ with $\mathcal{K} = 1.26 \cdot 10^{38}$ \rm erg s$^{-1}$): the value of $\lambda$ decreases with the BH spin until $a \sim 0.95$ (this value depends on the value of the viewing angle) where it reaches a minimum and then increases with $a$. We point out that this effect is only due to the fact that, for the same spectral peak, mass and accretion rate (as a function of the BH spin) change in a different way for different viewing angles.\footnote{For $a \lsim 0.95$, the value of $\lambda$ decreases as $a$ increases because the Eddington luminosity $L_{\rm Edd}$ increases by a factor of $\sim 2$ with respect to the disk luminosity $L_{\rm d}$, which is almost constant (there is a small increment by a factor of $\sim 1.2$); instead, for spin values close to the maximum, the disk luminosity increases more significantly than the Eddington luminosity due to a larger radiative efficiency and, for this reason, the value of $\lambda$ increases (and hence the minimum). For larger viewing angles ($\theta_{\rm v} > 50^{\circ}$) the value of $\lambda$ is always decreasing for all spin values.} We found the following upper limits: $\lambda < 0.5$ for J1342, $\lambda < 0.3$ for J1120, and $\lambda < 0.45$ for J0038, considering $\theta_{\rm v} < 45^{\circ}$ and spin $a > 0$ (Table \ref{ap:alltabresres}). Our KERRBB Eddington ratio estimates are smaller than the results found by BN18 for J1342 ($\lambda = 1.5^{+0.5}_{-0.4}$), MR11 for J1120 ($\lambda = 1.2^{+0.6}_{-0.5}$) and W18 for J0038 ($\lambda = 1.25 \pm 0.19$): the reason behind such different results is that they use bolometric luminosities (larger than our disk luminosity estimates) which lead to a larger value of $\lambda$, at least by a factor of $\sim 2$.

\subsection{KERRBB versus SLIMBH model}\label{2.5}

On the same Fig. \ref{KERRBBmass4}, we compared the results with the thin disk limit for the Eddington ratio following \citet{LaoNet} (black dotted line): in order to have a geometrically thin disk (i.e. ratio between the disk half-thickness and disk radial coordinate $z/r < 0.1$), the Eddington ratio must be $\lambda \lsim 0.3$ (see also \citealt{McClint}).\footnote{Only for J1342 and J0038 our estimates are above this limit.} Above this value, the results are not consistent with the thin disk assumptions and another solution (e.g. slim disk) must be used. In fact, beyond a critical accretion rate value, advection dominates over radiative energy transport and the disk becomes "slim". Also, the efficiency $\eta$ is lower because part of the energy dissipated in the disk is trapped in the accreting flow and not radiated away (e.g. \citealt{Katz}; \citealt{Begel}) and this effect is more prominent for large spin values (see \citealt{Sad09}). To fit the spectrum and infer physical quantities, we therefore used SLIMBH, a relativistic model that describes the emission produced by a slim disk which is thought to be more appropriate for bright AGNs (\citealt{Kora}): high redshift SMBHs (with possible disk accretion) may have gone through a fast accretion phase with a large $\dot{M}$. SLIMBH is designed for stellar and intermediate mass BHs: following the same analysis done by C18 for KERRBB, we found analytical expressions to describe the observed disk emission as a function of the BH mass, spin, Eddington ratio and viewing angle, in order to extend its usage also for SMBHs (see Appendix \ref{APP.B}). Figures \ref{KERRBBmass} and \ref{KERRBBmass4} show respectively the BH mass and the Eddington ratio for the three sources, computed using SLIMBH (dashed lines), for different spin values and angles, assuming the same peak position found with KERRBB: the results, summarized in Table \ref{ap:alltabresres}, are similar for low spin values and slightly different for large spin values (by a factor of $\sim 1.2 - 1.3$). Also for what concerns the observed disk luminosity and the BH spin, the results inferred from the SLIMBH fits are similar to those obtained using KERRBB.


\begin{figure*}
\centering
\hskip -0.2 cm
\includegraphics[width=0.53\textwidth]{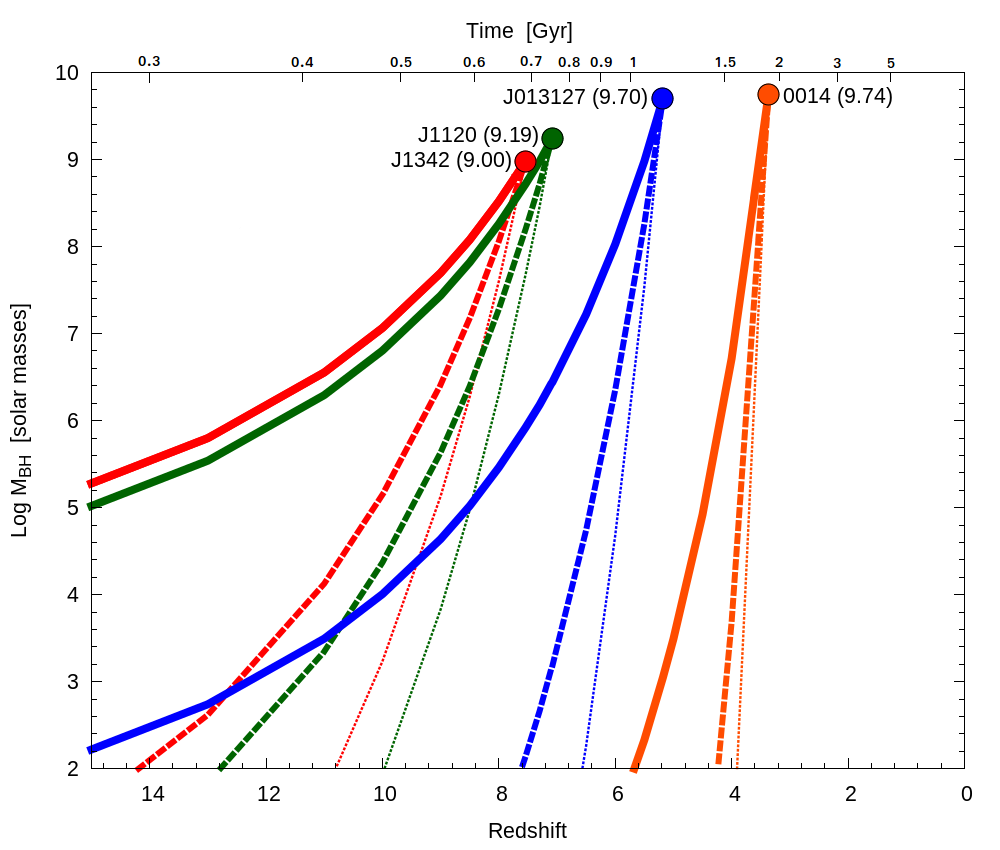} \includegraphics[width=0.47\textwidth]{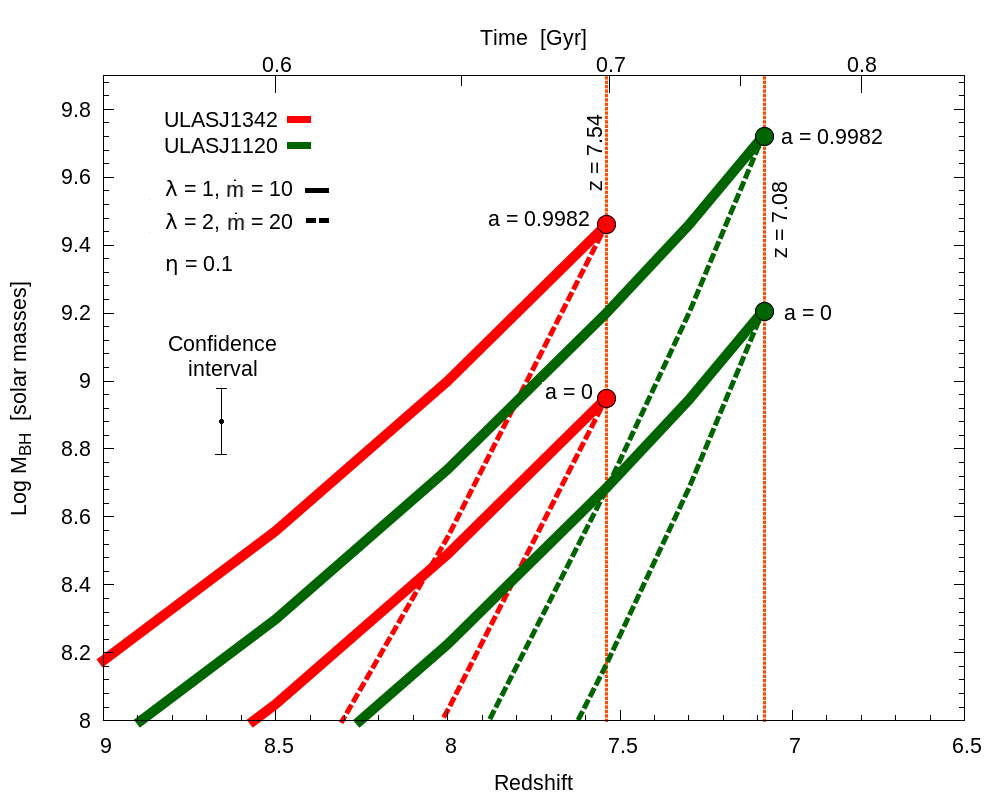}
\caption{Left panel: Black hole mass as a function of redshift (see Eq. \ref{Masstime}) for J1342 (red lines) and J1120 (green lines; these curves are related also to the source J0038 since the BH mass is similar to that of J1120). The evolutionary tracks are compared with the ones related to the sources SDSSJ013127.34-032100.1 (blue lines) and S5 0014+813 (orange lines). The final masses in the brackets correspond to the non-spinning BH computed for $\theta_{\rm v} = 30^{\circ}$ (for the case with $a = 0.9982$, the curves are similar and shifted rigidly to larger masses by a factor of $\sim 3.5$). Solid lines refer to $\lambda = 1$, dashed to $\lambda = 2$, dotted to $\lambda = 3$; the radiative efficiency $\eta = 0.1$ is fixed: for J1342, J1120 and J0038, in the case with $\lambda = 1$ ($\dot{m} = 10$) a massive seed of $\sim 10^4 M_{\odot}$ is required at redshift $z > 15$ in order to reach the observed mass. In the case of $\lambda = 2$, the evolution could have begun with a seed of $\sim 10^2 M_{\odot}$ at redshift $z \sim 13 - 14$, growing exponentially in $\sim 0.4$ Gyr. For $\lambda = 3$, a faster evolution of $\sim 0.2$ Gyr could have begun at $z \sim 10-11$. Instead for J013127 and 0014, if $\lambda \sim 1$, a seed of $\sim 10^2 M_{\odot}$ could have grown in $\sim 0.9$ Gyr, starting at redshift $z \sim 15$ and $z \sim 6$, respectively. In the case with a fixed Eddington ratio $\lambda = \eta\ \dot{m} \sim 1$, for the same value of $\dot{m}$, the radiative efficiency is smaller and the evolution is faster but the curves (not overplotted for clarity) show a similar trend. Right panel: zoom on the observed redshifts of J1342 (red lines) and J1120 (green lines) with the evolutionary tracks leading to black hole masses corresponding to the cases with $a=0$ and $a=0.9982$ (for J0038, the evolutionary tracks are similar to those of J1120, shifted to lower redshifts by $0.06$, not overplotted for clarity). Solid lines describe the case with $\lambda = 1$ and dashed lines the case with $\lambda = 2$. At redshift $z=7.54$, J1120 (J0038) could have been more (or less) massive than J1342 depending on the values of $\lambda$ and the BH spin: for example, in the case with $\lambda = 2$, J1120 (J0038) was less massive than J1342 by a factor $\gsim 2$ for any spin value; with $\lambda = 1$, J1120 (J0038) was more massive than J1342 only if the sources had different spins.}\label{EVOLUZ_L103}
\end{figure*}

\section{Evolution of the black hole mass} \label{sec:3}

In this section we describe the BH growth following different evolutionary scenarios, considering simplified assumptions discussed in the following. The aim of the estimates below is to explore possible evolutionary histories for the considered sources, so that they reach their masses via accretion at their observed (high) redshifts. We evaluate the evolutionary paths in terms of accretion parameters, radiative efficiency, seed mass and formation redshift. The procedure is oversimplified as no physically motivated time evolution in the accretion parameters is adopted. Nevertheless, despite the limitations of the procedure, it gives a sense of which range of evolutionary parameters is viable and which source provides the strongest constraints on the growth parameters.  

Accretion onto a BH changes its mass and spin (\citealt{Bardeen}; \citealt{Thorn74}). The variation of the BH mass $M$ as a function of time is
\[
	\frac{dM}{dt} = (1 - \eta) \dot{M},
\]

\noindent where $\eta$ is the radiative efficiency (depending on the BH spin) and $\dot{M}$ is the accretion rate. By integrating this expression, it is possible to find the evolution time $t_{\rm ev}$ for a BH, that is, the time it takes to grow from an initial mass $M_{\rm 0}$ to a final mass $M$ (Salpeter time, \citealt{Salpeter}) assuming a fixed Eddington ratio $\lambda$,
\begin{equation}\label{Masstime}
	\frac{t_{\rm ev}}{\rm{Gyr}} = 0.451 \frac{\eta}{1 - \eta} \frac{1}{\lambda} \ln \Bigl[ \frac{M}{M_{\rm 0}} \Bigl].
\end{equation}

\citet{Lapietal14} present a consistent scenario in which at high redshifts, a BH grows in a self-regulated regime with a radiative power slightly super-Eddington ($\lambda \lsim 4$) and a radiative efficiency $\eta = 0.15$. After this phase, a fast decrease of $\lambda$ occurs until the matter reservoir, assumed to surround the BH and accrete onto it, is exhausted (i.e. in a sub-Eddington phase). Figure 20 in \citet{Lapietal14} shows that the fast decrease of the Eddington ratio lasts less than $< 0.1$ Gyr while most of the BH evolution occurs during the super-Eddington phase (which lasts longer). Therefore, the BH evolution time can be approximated by Eq. \ref{Masstime} with $\lambda \geq 1$. Within such a scenario, our estimates for the Eddington ratios ($\lambda < 0.5$) would indicate that the three BHs we consider have already reached the last phase of their evolution.  Therefore, in order to asses their growth, we have to focus on the previous (super)-Eddington phase. 

A first oversimplified assumption is that the observed Eddington ratio and observed accretion rate (defined as $\dot{m} = \dot{M}/\dot{M}_{\rm Edd}$ where $\dot{M}_{\rm Edd} = L_{\rm Edd}/c^2$) describe the whole BH evolution. In this case, since for spin values $a>0$ we have inferred an Eddington ratio $\lambda < 1$ and an accretion rate $\dot{m} < 10$, a very massive seed (Log $M/M_{\odot} > 6.5$) is required for the sources at redshift $z \sim 40$. In other words, if the BH seed had a mass of $10^{2-4} M_{\odot}$, then a significantly larger $\lambda$ and $\dot{m}$ were needed during the fast growth phase of super-Eddington accretion. \footnote{For J1342, the evolution described with the observed parameters $\lambda \sim 0.56$ and $\dot{m} \sim 10$ is similar to the case with $\lambda \sim 1$ and $\dot{m} \sim 10$.}. Figure \ref{EVOLUZ_L103} shows the evolutionary tracks (i.e. BH mass as a function of redshift with the final mass corresponding to the case with $a=0$ and $\theta_{\rm v} = 30^{\circ}$) of the three sources, for different assumptions on the accretion parameters, as described below.\footnote{For the case where $a=0.9982$, the curves are similar and shifted rigidly to larger masses by a factor of $\sim 3.5$.} These are compared with the equivalent tracks for the radio loud sources SDSSJ013127.34-032100.1 (hereafter J013127, $z=5.18$) and S5 0014+813 (hereafter 0014, $z = 3.36$), studied by C18 using KERRBB, whose masses for non-rotating BHs are Log $M/M_{\odot} = 9.70$ and Log $M/M_{\odot} = 9.74$, respectively. For simplicity, we can describe the evolution in two ways:
\begin{enumerate}
	\item Super-Eddington evolution $\lambda \gsim 1$: We fixed the radiative efficiency to $\eta = 0.1$. We assume that this choice is reasonable both in the accretion disk scenario and the chaotic one. If we assume the presence of a slim disk in the super-Eddington phase, the BH can spin up to the maximum value ($a \sim 1$) and the radiative efficiency is lower than the canonical value $\sim 0.3$ (\citealt{Thorn74}) due to photon trapping; if instead the super-Eddington phase is characterized by a chaotic accretion, the BH spin is thought to be low or $\sim 0$ (\citealt{Dottietal}): the value of $\eta$ is not easy to estimated and we follow \citealt{Lapietal14}. In the range of $\dot{m}$ considered in this work, the choice of $\eta=0.1$ is in agreement with the results of \citet{Sad09} concerning the super-Eddington accretion through a slim disk. In the case $\lambda = 1$ ($\dot{m} = 10$), a massive seed of $\sim 10^4 M_{\odot}$ is required at redshift $z \sim 25 - 30$ or larger in order to reach the observed masses (as also obtained by BN18). For J013127 and 0014 instead, a seed of $\sim 10^2 M_{\odot}$ could have grown in $\sim 0.9$ Gyr, starting at redshift $z \sim 15$ and $z \sim 6$, respectively. For larger values of $\lambda$, the BH evolution could have begun with a lighter BH seed at smaller redshifts, growing exponentially and reaching the estimated masses in less than $\sim 0.4$ Gyr.	
	\item Eddington limited evolution $\lambda \sim 1$: In this case, by definition, we have the bound $\eta\ \dot{m} \sim 1$. For different values of $\dot{m}$, the evolutionary curves are similar to the ones in Fig. \ref{EVOLUZ_L103}. For the same value of $\dot{m}$, the evolution is slightly faster because $\eta$ is smaller due to the bound $\lambda \sim \eta\ \dot{m}$: assuming $\dot{m} = 10$, the radiative efficiency is $\eta = 0.1$ and this case is represented by the solid lines in Fig. \ref{EVOLUZ_L103}; for $\dot{m} = 20\ (30)$, we have $\eta = 0.05\ (0.03)$ and the evolution curves are similar to the ones described by dashed lines but the growth proceeds slightly faster due to an efficiency $< 0.1$.
\end{enumerate}

The right panel of Fig. \ref{EVOLUZ_L103} shows a zoom around the observed redshifts of the sources J1342 and J1120 (for J0038, the evolutionary tracks are similar to the ones of J1120, shifted to lower redshifts by $0.06$, not overplotted for clarity). At redshift $z=7.54$, J1120 and J0038 could have been more (or less) massive than J1342 depending on the values of the Eddington ratio and the BH spin: for example, in the case with $\lambda = 2$, J1120 and J0038 were less massive than J1342 by a factor of $\gsim 2$ for any spin value; in the case with $\lambda = 1$, the two sources were more massive than J1342 only if the sources had different spins. So, the source J1342 can be considered more extreme than J1120 and J0038, being more demanding to be built at such high redshift, in terms of seed mass or accretion speed.


\section{Discussion and conclusions} \label{sec-concl}

We adopted the relativistic thin disk model KERRBB and slim disk model SLIMBH to describe the optical-UV SED of the most distant quasars identified up to the end of 2018, ULASJ134208.10+092838.61 ($z=7.54$), ULASJ112001.48+064124.3 ($z = 7.08$) and DELSJ003836.10-152723.6 ($z=7.02$). The aim was to estimate their BH masses, spin and accretion rate. For the SED fitting process, we assumed a viewing angle $\theta_{\rm v} < 45^{\circ}$ (i.e. we assumed that $\theta_{\rm v}$ is smaller than the aperture angle of the molecular torus surrounding the accretion disk), consistent with the properties of a Type 1 QSO. The results of the modeling can be summarized as follows:
\begin{itemize}[label=$\bullet$]
	\item The masses of the three sources from the fit of the accretion disk emission are compatible with the virial masses estimated by \citet{Mor11}, \citet{Derosa},\citet{Banados17} and \citet{Wangetal18}. Our uncertainties are smaller compared to the systematics of virial estimates ($\sim 0.5$ dex); 
	\item On the assumption that the virial estimates are reliable measurements of the BH masses (with no systematic uncertainties involved), our findings imply that for the source J1342 the BH spin has to be $a < 0.5$. For J1120, an upper limit for the BH spin is present only for viewing angles $\theta_{\rm v} > 0^{\circ}$ ($a<0.9$ for $\theta_{\rm v} = 30^{\circ}$); for J0038 the BH spin has to be $a < 0.3$;
	\item The three sources emit radiation at a sub-Eddington rate, contrary to what \citet{Mor11}, \citet{Banados17} and \citet{Wangetal18} found in their work: they computed the Eddington ratio $\lambda$ using the AGN bolometric luminosity, which includes the IR and X-ray contributions, resulting in larger Eddington ratios at least by a factor of $\sim 2$. Our results are in agreement with several works (e.g. \citealt{Wiletal}) showing that at high redshifts not all the sources emit close to the Eddington limit. Clearly a larger sample is needed to further investigate this issue. The sub-Eddington regime we inferred for the observed accretion suggests the need of a previous (super-Eddington?) phase during which most of the BH mass was quickly assembled (e.g. \citealt{Lapietal14});		
	\item We showed, with a simplified approach, possible paths for the BH growth evolution. Two scenarios were considered: [1] a super-Eddington evolution with $\lambda > 1$; [2] an Eddington limited evolution with $\lambda \sim 1$. For both these scenarios, we considered different accretion rates $\dot{m}$ and different radiative efficiencies $\eta$ (with the bound $\lambda = \eta\ \dot{m}$). The evolutionary tracks (i.e. mass versus time) of the three sources were also compared with those of SDSSJ013127.34-032100.1 ($z=5.18$) and S5 0014+813 ($z = 3.36$), studied by \citet{Campiti}. Independently of the value of $\lambda$, we found that a high accretion rate ($\dot{m} \sim 15-30$) and a low radiative efficiency ($\sim 10 \%$) are necessary to reach the estimated masses at the observed redshifts, when starting the evolution from a $\sim 10^{2-4} M_{\odot}$ black hole seed at redshift $z \sim 10-20$. It was also shown that the source J1342 can be considered more extreme than J1120 and J0038, in the sense of being more demanding to be built at such high redshift in terms of seed mass or accretion speed.
\end{itemize}

\noindent Some caveats and issues should be mentioned regarding our analysis:
\begin{itemize}[label=$\bullet$]
	\item We checked the robustness of our results, that is, the physical parameters derived from the model fits against emission and absorption affecting the observed spectra. Even for the most significant effect which can be due to (low level) dust absorption, our results (disk luminosity, BH mass, Eddington ratio) differ at most by a factor of $\lsim 2$. In the absence of dust absorption the main uncertainties on $L^{\rm obs}_{\rm d}$, $M$ and $\lambda_{\rm Edd}$ are limited by the confidence interval of the spectral modeling (`visually' represented by the red curves in Fig. \ref{SED}); 
	\item In the (super-) Eddington phase of the assumed BH evolution, we considered constant parameters ($\lambda$, $\eta$, $\dot{m}$). This is clearly an oversimplified approach due to our ignorance of the temporal and physical evolution of the parameters regulating the growth history. The presented growth tracks have to be considered only as an indicator of the parameter space involved in order to account the measured mass. We note that, by definition, the assumption $\dot{m} \sim constant$ means that the accretion rate increases with the BH mass;
	\item On the same line, we did not account for any kind of transitions between the (super-) Eddington phase and the observed sub-Eddington one; 
	\item Similarly, evolution of the BH spin has been neglected as this cannot be simply predicted and it would depend on the accretion rate and modality. One possible solution for the evolution is given by a two-phase scenario:
 		\begin{itemize}
			\item chaotic growth of the BH via BH-BH mergings and/or accretion of matter and gas falling onto a seed BH with no/small net angular momentum ($a \sim 0$ or low would be expected). The low efficiency $\eta \lsim 0.1$ could have provided a rapid growth of the BH even without violating the Eddington limit;
			\item stable phase in which the accretion occurs through a disk-like structure (i.e. the phase we observe) producing the optical-UV bump: this kind of accretion could spin the BH up to its maximum rotation.
	\end{itemize}
\end{itemize}

Future works on a large sample of quasars at $z > 7$ could shed light on the possible mass and Eddington ratio distributions and give stronger constraints on the probable BH growth. The usage of relativistic accretion disk models (e.g. KERRBB and SLIMBH) could be a viable, alternative (and possibly more accurate) means to infer parameters like BH mass and Eddington ratio. In principle, well-calibrated alternative mass determinations would constrain BH spins. \\


\begin{acknowledgements}
We are grateful to the anonymous referee for her/his constructive critical comments and suggestions, which helped improving the paper.
\end{acknowledgements}


\medskip
\let\itshape\upshape

\begingroup
\let\clearpage\relax


\appendix

\section{KERRBB equations} \label{APP.A}

\begin{table*} 
\centering
\footnotesize
\begin{tabular}{lllllllllll}
\hline 
\vspace{1mm}
$\theta_{\rm v}$ & & $\alpha_{\rm g}$ & $\beta_{\rm g}$ & $\gamma_{\rm g}$ & $\delta_{\rm g}$ & $\epsilon_{\rm g}$ & $\zeta_{\rm g}$ & $\iota_{\rm g}$ & $\kappa_{\rm g}$ \\ 
\hline 
\hline\\
$30^{\circ}$ & $g_{\rm 1}$ & -6238.24278 & -0.12968 & -926.60788 & 37153.52731& -319409.16085 & 983173.49265 & -1228407.5946 & 532213.69610 \\
\vspace{2mm}
& $g_{\rm 2}$ & -9332.65475 & -0.30891 & -1356.5323 & 54978.79312 & -474342.11691 & 1462740.0025 & -1829629.6737 & 793292.21583 \\
$45^{\circ}$ & $g_{\rm 1}$ & -12431.6125 & -0.20298 & -1797.0174 &  73606.04261 & -636017.88764 & 1961099.1802 & -2451848.7325 & 1062521.5524 \\
& $g_{\rm 2}$ & -16314.3127 & -0.41098 & -2325.9424 &  95651.40409 & -828227.37848 &  2557278.0153 & -3200363.0119 & 1387917.3481 \\
\hline \\
\end{tabular}
\caption{KERRBB parameter values of Eq. \ref{eq:functional22} for the equations $g_{\rm 1}$ and $g_{\rm 2}$, for the viewing angles $\theta_{\rm v} = 30^{\circ} - 45^{\circ}$.\label{ap:tab1}}
\end{table*}

\begin{table*} 
\centering
\footnotesize
\begin{tabular}{lllllllllll}
\hline\\
\vspace{1mm}
$\theta_{\rm v}$ & & $\bar{a}$ & $\bar{b}$ & $\bar{c}$ & $\bar{d}$ & $\bar{e}$ & $\bar{f}$ \\
\hline 
\hline \\
$0^{\circ}$ & $\alpha_{\rm 1,s}$ & 1.37528 & -0.03581 & 0.21458 & -0.35971 & 0.16821 & 0\\
& $\beta_{\rm 1,s}$ & -0.69241 & 0.21439 & -0.57440 & 0.75878 & -0.31345 & 0 \\
& $\gamma_{\rm 1,s}$ & -0.28949 & 0.33854 & -1.10213 & 1.27581 & -0.48471 & 0\\
& $\delta_{\rm 1,s}$ & -0.02094 & 0.20299 & -0.65531 & 0.69134 & -0.24475 & 0\\
& $\epsilon_{\rm 1,s}$ & 0.00538 & 0.04041 & -0.12718 & 0.12852 & -0.04411 & 0\\
\hline\\
& $\alpha_{\rm 2,s}$ & 9.917034 & -3.22512 & 17.89987 & -41.84139 & 41.27971 & -14.78658 \\
& $\beta_{\rm 2,s}$ & 3.51987 & -2.44127 & 13.36609 & -29.61009 & 29.12455 & -10.46037 \\
& $\gamma_{\rm 2,s}$ & 0.72886 & -4.09764 & 21.16717 & -46.66724 & 46.53742 & -16.81675 \\
& $\delta_{\rm 2,s}$ & 0.39809 & -2.70631 & 14.15261 & -31.93365 & 32.12041 & -11.63591 \\
& $\epsilon_{\rm 2,s}$ & 0.10769 & -0.52470 & 2.79254 & -6.40072 & 6.46561 & -2.34359 \\
\hline
\hline \\
$30^{\circ}$ & $\alpha_{\rm 1,s}$ & 1.43631 & -0.04881 & 0.29277 & -0.47116 & 0.22691 & 0\\
& $\beta_{\rm 1,s}$ & -0.79532 & -0.01997 &  0.33520 & -0.75650 & 0.53158 & 0 \\
& $\gamma_{\rm 1,s}$ & -0.22587 & -0.02063 &  0.09107 & -0.69291 & 0.63651 & 0\\
& $\delta_{\rm 1,s}$ & 0.02389 & 0.02883 & -0.15739 & -0.10901 & 0.22802 & 0\\
& $\epsilon_{\rm 1,s}$ & 0.01253 & 0.01160 & -0.05570 & 0.01933 & 0.02243 & 0\\
\hline\\
& $\alpha_{\rm 2,s}$ & 10.35255 & -3.29155 & 18.23667 & -42.38302 & 41.68164 & -14.90491 \\
& $\beta_{\rm 2,s}$ & 2.98930 & -2.42948 & 12.92896 & -28.15866 &  27.60484 & -9.89120 \\
& $\gamma_{\rm 2,s}$ & 0.59669 & -4.53087 &  22.72740 & -49.83698 & 49.55957 & -17.85179 \\
& $\delta_{\rm 2,s}$ & 0.48252 & -3.02872 &  15.50716 & -34.91153 &  35.02577 & -12.65587 \\
& $\epsilon_{\rm 2,s}$ & 0.13420 & -0.58612 &  3.06804 & -7.02543 &  7.08401 & -2.56372 \\
\hline
\hline \\
$45^{\circ}$ & $\alpha_{\rm 1,s}$ & 1.50731 & -0.07421 & 0.44310 & -0.66183 & 0.30977 & 0 \\
& $\beta_{\rm 1,s}$ & -0.94113 & 0.04156 & 0.22037 & -0.66194 & 0.48374 & 0 \\
& $\gamma_{\rm 1,s}$ & -0.13063 & -0.33330 & 1.96527 & -3.88455 & 2.23797 & 0 \\
& $\delta_{\rm 1,s}$ & 0.06519 & -0.16637 &  1.16515 & -2.43971 & 1.43194 & 0 \\
& $\epsilon_{\rm 1,s}$ & 0.01540 & -0.02046 &  0.19603 & -0.44262 & 0.26763 & 0\\ 
\hline\\
& $\alpha_{\rm 2,s}$ & 11.07996 & -3.55277 & 19.73825 & -45.64546 & 44.84619 & -16.02576 \\
& $\beta_{\rm 2,s}$ & 2.34516 & -2.75232 & 14.20309 & -30.41746 & 29.43857 & -10.38000 \\
& $\gamma_{\rm 2,s}$ & 0.65369 & -5.59958 & 27.68371 & -60.66151 & 59.68996 & -21.22119 \\
& $\delta_{\rm 2,s}$ & 0.71299 & -3.70302 & 18.84159 & -42.43550 & 42.23110 & -15.11877 \\
& $\epsilon_{\rm 2,s}$ & 0.18732 & -0.70811 & 3.69829 & -8.47710 & 8.49652 & -3.05447 \\
\hline
\hline \\
\end{tabular}
\caption{SLIMBH parameter values of Eq. \ref{poly} for the viewing angles $\theta_{\rm v} = 0^{\circ} - 30^{\circ} - 45^{\circ}$. The subscript $\rm i = 1,2$ specifies the equations $g_{\rm 1,s}$ and $g_{\rm 2,s}$ to which the parameters are related. \label{ap:tab1sl}}
\end{table*}

The relativistic model KERRBB (\citealt{Lietal}) describes the emission produced by a thin disk around a Kerr BH. Using a ray-tracing technique to compute the observed spectrum, the authors included all relativistic effects such as frame-dragging, Doppler beaming and light-bending. It is an extension of a previous relativistic model called GRAD (\citealt{Hanawa}; \citealt{Ebietal}) which assumes a non-rotating black hole. \citet{Campiti} build an analytic approximation of the KERRBB disk emission features considering an hardening factor $f = 1$ and no limb-darkening effect: in the case of a face-on disk, they found analytic expressions to compute the BH mass and accretion rate by fitting a given SED for different spin values. Here we followed the same procedure. From the SED, the spectrum peak $\rm \nu_p$ and luminosity $\rm \nu_p L_{\nu_p}$ are
\begin{equation} \label{eq:nupeak2}
	\frac{\nu_{\rm p}}{\text{[Hz]}} = \mathcal{A} \left[ \frac{\dot{M}}{M_{\odot} \text{yr}^{-1}} \right]^{1/4} 
\left[\frac{M}{10^9 M_{\odot}} \right]^{-1/2} g_{\rm 1}(a, \theta_{\rm v}),
\end{equation}
\begin{equation} \label{eq:nulnupeak2}
	\frac{\rm \nu_{\rm p} L_{\nu_{\rm p}}}{\text{[erg/s]}} = \mathcal{B} \left[ \frac{\dot{M}}{M_{\odot} \text{yr}^{-1}} \right] \cos \theta_v\ g_{\rm 2}(a, \theta_{\rm v}),
\end{equation}
	
\noindent where Log $\mathcal{A} = 15.25$, Log $\mathcal{B} = 45.66$. The functions $g_1$ and $g_2$ account for all the effects due to the viewing angle $\theta_{\rm v}$ and the spin $a$. We fixed $\theta_{\rm v}$ and we used KERRBB data to find analytic expressions for $g_{\rm 1}$ and $g_{\rm 1}$ that can be written as
\begin{eqnarray} \label{eq:functional22}
	g_{\rm i}(a, \theta) &=& \alpha_{\rm g} + \beta_{\rm g} y_{\rm 1} + \gamma_{\rm g} y_{\rm 2} + \delta_{\rm g} y_{\rm 3} + \epsilon_{\rm g} y_{\rm 4} +\zeta_{\rm g} y_{\rm 5} + \iota_{\rm g} y_{\rm 6} + \kappa_{\rm g} y_{\rm 7}
	\nonumber \\
	y_{\rm n} &\equiv& \log(n-a) \qquad \qquad \qquad \rm i = 1, 2
\end{eqnarray}
	
A smaller number of parameters reduces the precision of these equations. The different parameters for $g_{\rm 1}$ and $g_{\rm 2}$ are reported in Table \ref{ap:tab1} for the viewing angles $\theta_{\rm v} = 30^{\circ} - 45^{\circ}$. Then the BH mass, accretion rate and Eddington ratio can be found:
\[
	\frac{M}{10^9 M_{\odot}} = \Bigl[ \frac{g_{\rm 1}(a, \theta_{\rm v}) \mathcal{A}}{\rm \nu_p}  \Bigl]^2 \sqrt{\frac{\rm \nu_p L_{\nu_p}}{\mathcal{B} \cos \theta_{\rm v}\ g_{\rm 2}(a, \theta_{\rm v})}},
\]
\[
	\frac{\dot{M}}{M_{\odot} \rm{yr}^{-1}} = \frac{\rm \nu_p L_{\nu_p}}{\mathcal{B} \cos \theta_{\rm v}\ g_{\rm 2}(a, \theta_{\rm v})},
\]
\[
	\lambda = \mathcal{D}\ \frac{\eta(a)}{g^2_{\rm 1}(a, \theta_{\rm v}) \sqrt{\cos \theta_{\rm v}\ g_{\rm 2}(a, \theta_{\rm v})}} \rm \nu^2_p \sqrt{\rm \nu_p L_{\nu_p}} ,
\]

\noindent where Log $\mathcal{D}=-53.675$. For $\theta_{\rm v} = 0^{\circ}$ we used the expressions and the parameters found by \citet{Campiti}.

	
\section{SLIMBH equations} \label{APP.B}

The relativistic model SLIMBH (\citealt{Abretal}; \citealt{Sad09}; \citealt{SadwAbra09}; \citealt{SadwAbra}; \citealt{StraubDo}) describes the emission produced by a slim disk around a BH. It is based on the relativistic description of \citet{NovTho} and it accounts also for the vertical radiative energy transport which is not negligible for high accretion rates. As KERRBB, the observed spectrum is computed using the ray-tracing technique and it is implemented in XPSEC. For this work, we assumed a viscosity with $\alpha = 0.1$, hardening factor $f = 1$, and no limb-darkening effect. The procedure to find analytic expression for SLIMBH is similar to the one adopted for KERRBB. Since the Eddington ratio is a free parameter of the model, we re-wrote Eqs. \ref{eq:nupeak2} and \ref{eq:nulnupeak2}, adding $\lambda$ and including all the effects due to spin and viewing angle in the new functions $g_{\rm 1,s}$ and $g_{\rm 2,s}$:
\begin{equation} \label{eq:nupeakslim}
	\frac{\rm \nu_{\rm p}}{\text{[Hz]}} = 1.22\ \mathcal{A}\ \lambda^{1/4} 
\left[\frac{M}{10^9 M_{\odot}} \right]^{-1/4} g_{\rm 1,s}(a, \theta_{\rm v}, \lambda),
\end{equation}
\begin{equation} \label{eq:nulnupeakslim}
	\frac{\rm \nu_{\rm p} L_{\nu_{\rm p}}}{\text{[erg/s]}} = 2.21\ \mathcal{B}\ \lambda \left[ \frac{M}{10^9 M_{\odot}} \right] \cos \theta_{\rm v}\ g_{\rm 2,s}(a, \theta_{\rm v}, \lambda).
\end{equation}

The new functions $g_{\rm 1,s}$ and $g_{\rm 2,s}$ depend also on the Eddington ratio. We adopted the following procedure: we fixed the viewing angle $\theta_{\rm v}$, the BH mass $M$ and the Eddington ratio $\lambda$, and we used SLIMBH to compute the peak frequency and luminosity for different spin values; then we found analytic equations for $g_{\rm 1,2}$ and $g_{\rm 2,s}$ and, after that, we considered the product
\[
	\rm [\nu_p L_{\nu_p}]^{1/4} \nu_p = \mathcal{E}\ \Bigl[ g_{2,s}(a, \theta_v, \lambda) \cos \theta_v \Bigl]^{1/4} \Bigl[ g_{1,s}(a, \theta_v, \lambda) \Bigl] \sqrt{\lambda},
\]

\noindent where Log $\mathcal{E} = 26.84$. At this point, we estimated the left-hand side of the expression from an observed spectrum. The right-hand side is derived since we have found analytic expressions for $g_{\rm 1,s}$ and $g_{\rm 2,s}$: the comparison leads to the only value of the BH spin corresponding to the fixed $\lambda$. By using this spin value in Eq. \ref{eq:nupeakslim} (or \ref{eq:nulnupeakslim}), it is possible to find the corresponding BH mass. We repeated this procedure for different Eddington ratio values and found the solutions for different BH spins. The procedure led to the following analytic functions for $g_{\rm 1,s}$ and $g_{\rm 2,s}$:
\begin{eqnarray} \label{eq:functional2}
	g_{\rm i,s}(a, \theta_{\rm v}, \lambda) &=& \alpha_{\rm i,s} + \beta_{\rm i,s} y_{\rm 1} + \gamma_{\rm i,s} (y_{\rm 1})^2 + \delta_{\rm i,s} (y_{\rm 1})^3 + \epsilon_{\rm i,s} (y_{\rm 1})^4
	\nonumber \\
	y_{\rm 1} &\equiv& \log(1-a)	 \qquad \qquad \qquad \rm i = 1,2
	\nonumber
\end{eqnarray}

\noindent The parameters $\alpha_{\rm i,s}$, $\beta_{\rm i,s}$, $\gamma_{\rm i,s}$, $\delta_{\rm i,s}$, $\epsilon_{\rm i,s}$ are a function of the Eddington ratio $\lambda$ and can be approximated with a polynomial
\begin{equation} \label{poly}
	\chi_{\rm i,s}(\theta_{\rm v}, \lambda) = \bar{a} + \bar{b} \lambda + \bar{c}\lambda^2 + \bar{d} \lambda^3 + \bar{e} \lambda^4 + \bar{f} \lambda^5
\end{equation}

\noindent whose parameter values are reported in Table \ref{ap:tab1sl}, for the viewing angles $\theta_{\rm v} = 0^{\circ} - 30^{\circ} - 45^{\circ}$. Therefore, as for KERRBB, only the spectrum peak position (frequency $\nu_{\rm p}$ and luminosity $\nu_{\rm p} L_{\nu_{\rm p}}$) are required in order to extrapolate information about the BH. 


\label{lastpage}
\endgroup

\end{document}